\documentclass[12pt,a4paper]{iopart}
\expandafter\let\csname equation*\endcsname\relax
\expandafter\let\csname endequation*\endcsname\relax
\usepackage{amsmath}
\usepackage{iopams}

\usepackage{bm}
\usepackage{upgreek}
\usepackage[utf8]{inputenc}
\usepackage{amsfonts}
\usepackage{amssymb}
\usepackage{graphics,graphicx}
\usepackage[T1]{fontenc}
\usepackage[english]{babel}
\usepackage{cite}
\usepackage{mathtools}
\usepackage{empheq}

\usepackage{upgreek}

\begin{document}
\title[Electromagnetic Field Distribution and Divergence-Dependence ...]{Electromagnetic Field Distribution and Divergence-Dependence
of a Radially Polarized Gaussian Vector Beam Focused by a
Parabolic Mirror}
\author{László Pálfalvi$^{1,2}$, Zerihun Tadele Godana$^{1,3}$* and János Hebling$^{1,3}$}
\address{$^1$ Institute of Physics, University of Pécs, H‐7624 Pécs, Hungary\\
$^2$ HUN-REN--PTE High‐Field Terahertz Research Group, H‐7624 Pécs, Hungary\\
$^3$ Szentágothai Research Centre, University of Pécs, H‐7624 Pécs, Hungary}
\ead{godanaze@gamma.ttk.pte.hu}

\vspace{10pt}
\begin{indented}
\item[]May 2024
\end{indented}

\begin{abstract}
In this work, we derived formulae concerning the electric and magnetic field characteristics of a focused radially polarized Gaussian vector beam. Such a beam is consistent with Maxwell's equations contrary to plane waves having uniform field distribution. Hence a realistic picture is provided of the focused field distributions having importance before designing applications such as particle acceleration. For focusing a perfectly reflecting large numerical aperture on-axis parabolic mirror was supposed to have practical importance. The computation technique was based on the Stratton-Chu vector diffraction method. We pointed out that this offers a unique opportunity in the long wavelength regime, where the Richards-Wolf theory becomes unreliable. In the terahertz frequency range longitudinal electric field component with an amplitude of $\sim$160 $\text{MV}/\text{cm}$ was predicted, which is ideal for particle acceleration applications. Based on the field characteristics experienced as a function of the focusing angle, the possibility of using a paraboloid ring for particle acceleration was suggested. Its advantage is reflected not only in the strong available longitudinal field but also in ensuring the unobstructed transfer of particles as a practical point of view. The axial and radial distributions of the longitudinal electric field component for different incident beam divergences were analyzed in detail. It was found that the shift of the physical focus relative to the geometrical focus along the longitudinal direction shows a linear dependence on the divergence. The effect of the divergence angle on the field enhancement factor was also studied. 

  \vspace{\the\baselineskip}


\end{abstract}

\maketitle

\vspace{5mm}
\section{Introduction}
\noindent

Applications of the longitudinal electric field component of a tightly focused intense electromagnetic field such as electron acceleration are current research areas \cite{Popov2008, Wong2010, Payeur2012, Wong2013}. For efficient acceleration high electric field component along the motion direction of the particle as well as a sufficiently large interaction length is required. Due to the absence of spherical aberration and the possibility of using them in a large numerical aperture configuration parabolic mirrors with high reflectivity are excellent candidates for focusing. They are typically used both in on- and off-axis configurations. A radially polarized incident beam is preferred to achieve a large longitudinal focused electric field instead of a linearly polarized one. Focusing radially polarized beams with a large numerical aperture parabolic mirror has been discussed in several works \cite{Lieb2001,Davidson2004,Bokor2008,Quabis2000,Youngworth2000,Lindlein2007,April2010,Dehez2012,Lindlein2021} based on the Richards-Wolf theory \cite{Richards}. These researches were mostly motivated by developments for microscopy \cite{Lieb2001,Debus2003,Stadler2008}. 

In the course of the particle acceleration developments knowing exactly the characteristics of the electromagnetic field of a focused beam in the focal region is also essential. The Stratton–Chu vector diffraction theory \cite{Stratton1939} based approximations often provide more accurate results than the Richards-Wolf based, especially in the long wavelength cases.  Focusing linearly polarized, monochromatic electromagnetic plane waves \cite{Török2000,Varga2000}, and pulses \cite{Zhang2023} by a paraboloid is already elaborated based on the Stratton–Chu theory. The discrepancy between the Richards–Wolf theory \cite{Richards} based results \cite{Sheppard1977} and the Stratton–Chu theory based results \cite{Török2000} found for the large numerical aperture cases was clarified in \cite{Lindlein2021}.      

Based on the Stratton–Chu integrals, recently, we presented a derivation giving the electric field, when a radially polarized, monochromatic beam with uniform amplitude is focused by an on-axis parabolic mirror \cite{Godana2023}. 

However, a radially polarized beam with uniform cross-sectional amplitude is inconsistent with Maxwell’s equations. For example, it is immediately apparent, that it contradicts Gauss’s Law. Starting from the paraxial wave equation concerning the vector potential in Lorentz gauge, Kirk T. McDonalds has found an \textit{Axicon Gaussian Vector Beam} solution having axial and radial electric, and azimuthal magnetic field components \cite{McDonald2000}. Rigorously, the requirement for the applicability of the Stratton-Chu integrals is that the beam going to be focused has to be consistent with Maxwell's equations.

In this paper, we derive formulae concerning the radial and axial electric and the azimuthal magnetic field components of a radially polarized vector Gaussian beam focused by a perfectly reflecting on-axis parabolic mirror having an arbitrary numerical aperture. The field characteristics are analyzed in the focal region for a particular case with a large numerical aperture. The possibility of using a paraboloid ring as a practical device for the technical implementation of the particle acceleration is raised and its performance in reaching a strong longitudinal field component is discussed. Since the introduced model can treat divergent beams as well, some issues will be addressed concerning it. The variation of the field distributions (among others the shift of the focus and the variation of the field enhancement factor) is analyzed versus the divergence angle. The advantages of the THz wavelength are emphasized, and particular examples will be given.  Although our investigation is motivated by particle acceleration applications, studying the acceleration mechanism itself is not the purpose of the present paper.

\section{The Radially Polarized Gaussian Beam}
\noindent

Linearly polarized laser beams usually behave according to the well-known Gaussian beam formula derived from the paraxial wave equation concerning the electric field \cite{Saleh2019, Guenther2015}. However, this is not the only solution that can be obtained from Maxwell’s equations by reasonable approximations. Kirk T. McDonalds has derived the formulae of a radially polarized Gaussian Vector Beam also known as an Axicon Beam \cite{McDonald2000}. His approach was based on a paraxial approximation applied in the wave equation concerning the vector potential.

In a cylindrical coordinate system using Gaussian units the radial, azimuthal, and axial electric and magnetic field components of a cw (monochromatic) Axicon Gaussian Beam propagating into the $-z$ direction are \cite{McDonald2000}.
\begin{equation} \label{eq1}
\begin{split}
&\mathcal{E}_\rho(\rho,z)=\mathcal{E}_0\dfrac{\rho}{w_0}\left[f(z)\right]^2\exp\left(-f(z)\dfrac{\rho^2}{w_0^2}-ikz\right),\\
 &\mathcal{E}_\phi=0,\\
&\mathcal{E}_z(\rho,z)=i\theta_0\mathcal{E}_0\left[f(z)\right]^2\left(1-f(z)\dfrac{\rho^2}{w_0^2}\right)\exp\left(-f(z)\dfrac{\rho^2}{w_0^2}-ikz\right)
\end{split}
\end{equation}
and
\begin{equation} \label{eq2}
\begin{split}
&\mathcal{H}_\rho=0, \\
&\mathcal{H}_\phi(\rho,z)=\mathcal{E}_\rho(\rho,z),\\
&\mathcal{H}_z=0
\end{split}
\end{equation}
apart from the \(e^{-i\omega t}\) factor, where
\begin{equation} \label{eq3}
f(z)=\dfrac{1}{1-i\dfrac{z}{z_0}}=\dfrac{1+i\dfrac{z}{z_0}}{1+\dfrac{z^2}{z_0^2}}=\dfrac{\exp\left[i\tan^{-1}\left(\dfrac{z}{z_0}\right)\right]}{\sqrt{1+\dfrac{z^2}{z_0^2}}},
\end{equation}
and \(w(z)=w_0\sqrt{1+\dfrac{z^2}{z_0^2}}\) is the beam radius at \(z\). \(w_0\) is the beam waist (at \(z=0\)), \(z_0\) is the Rayleigh range, and \(\theta_0\) is the diffraction angle which are related by
\begin{equation} \label{eq4}
\theta_0=\dfrac{w_0}{z_0} \quad \textrm{and} \quad z_0=\dfrac{kw_0^2}{2}.
\end{equation}
In Eqs. \ (\ref{eq1}) and (\ref{eq2}) the terms proportional to the second and higher order powers of \(\theta_0\) are neglected.

\begin{figure}[h]
\centering
\includegraphics[width=10.5 cm]
{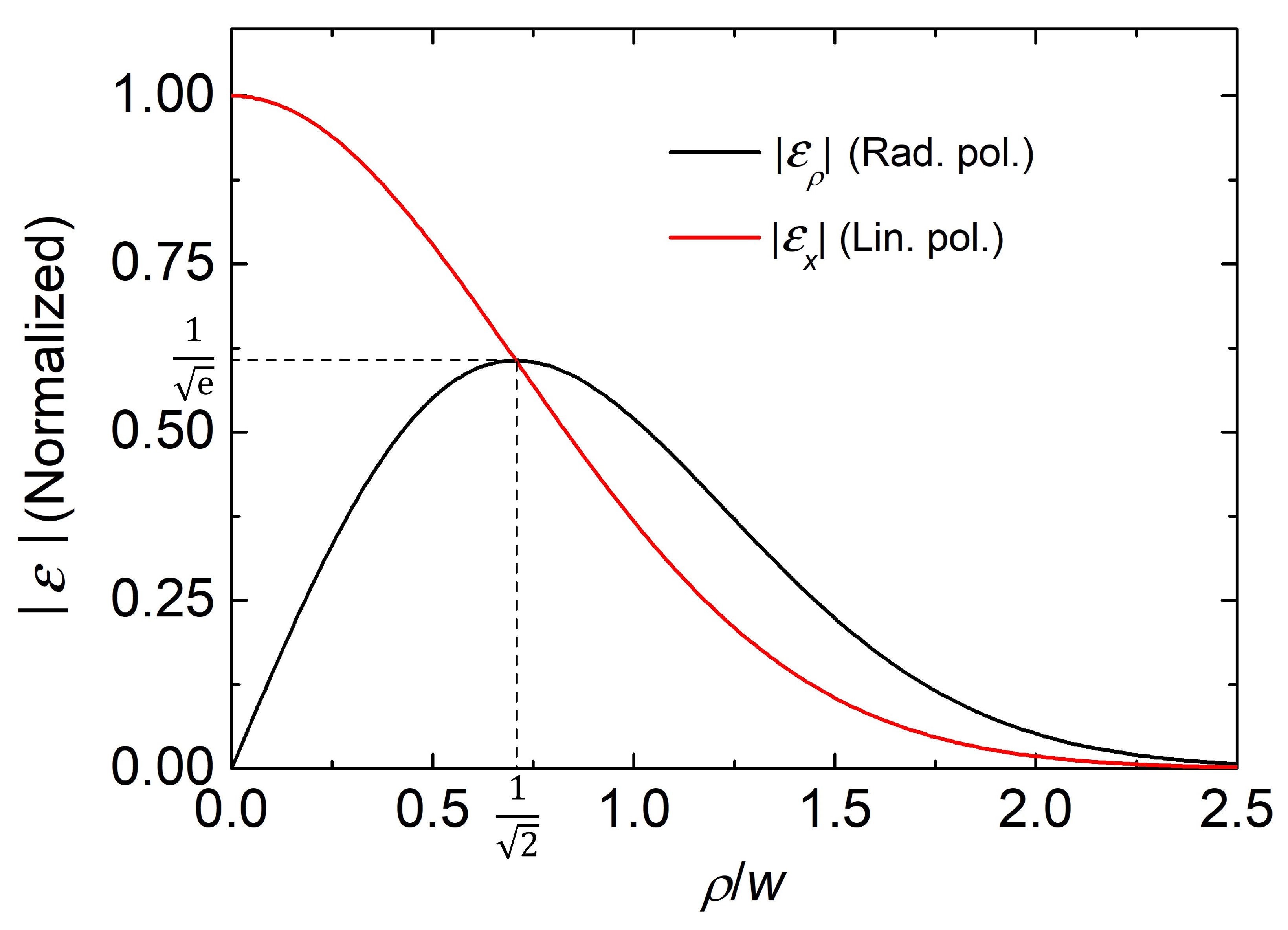}
\caption{The \(|\mathcal{E}_{\rho}|\) (\(|\mathcal{E}_x|\))  amplitude of the transversal electric field component
of a radially (black line) and a linearly polarized (red line) Gaussian beam.\label{fig1}}
\end{figure}

In Fig. \ \ref{fig1} the radially and linearly polarized Gaussian beams are compared based on the radial distribution of the amplitude of their transversal electric field component \(|\mathcal{E}_\rho|\) and \(|\mathcal{E}_x|\), respectively, where
\begin{equation} \label{Ex}
\mathcal{E}_x(\rho,z)=\mathcal{E}_0\cdot f(z)\cdot \exp\left(-f(z)\dfrac{\rho^2}{w_0^2}-ikz\right).
\end{equation}
The \(|\mathcal{E}_\rho|\) of a radially (see Eq.\ (\ref{eq1})) and the \(|\mathcal{E}_x|\) of a linearly (see Eq.\ (\ref{Ex}))
polarized Gaussian beam are plotted in Fig. \ \ref{fig1} versus \(\rho/w\). \(|\mathcal{E}_\rho|\) was normalized to
1. The power of the two beams was assumed to be equal as a normalizing condition for \(|\mathcal{E}_x|\). Note that while \(w\) is regarded as the characteristic size of the linearly polarized beam (where \(|\mathcal{E}_x|\) reaches \(1/e\) of its maxima), \(w/\sqrt{2}\) is the characteristic size of the radially polarized beam, where \(|\mathcal{E}_\rho|\) reaches its maxima. At \(\rho/w=1/\sqrt{2}\) both electric field amplitude equals \(1/\sqrt{e}\). Owing to the characteristics of the transversal distribution of \(|\mathcal{E}_\rho|\) shown in the figure, such beams are called \textit{Doughnut} Beams. Note, that the curves shown in Fig. \ \ref{fig1} are independent of the \(z\) position. Therefore, \(w\) was used, and in the Figure without any argument above.

\section{Expressions for \(\Vec{E}\) and \(\Vec{H}\), when a Radially Polarized Gaussian Beam is focused by a Parabolic Mirror}

\noindent

For fields oscillating with \(\omega\) angular frequency, Stratton and Chu constituted a formula pair for the electric and magnetic fields, regarded as the basic equations of the vector diffraction theory \cite{Stratton1939}. Although these formulae have already been known since 1939, they are even today frequently used for the solution of various basic diffraction problems \cite{Török2000, Hsu1994}, as well as for practical applications \cite{Phan2019, Grand2020, Guerboukha2020}, especially for beam focusing in a large numerical aperture geometry \cite{Török2000, Varga2000, Zhang2023, Godana2023}.

The Stratton–Chu formulae refer to a discontinuous (open) surface \textit{S}, bounded by a closed contour \textit{C}. They read as
\begin{eqnarray}
\label{eq5}
\vec{E}(\vec{r})=
&\dfrac{1}{4\pi}&\int_{S}\left[ik\left(\vec{n}\times\vec{\mathcal{H}}\right)G
+\left(\vec{n}\times\vec{\mathcal{E}}\right)\times\vec{\nabla} G
+\left(\vec{n}\cdot\vec{\mathcal{E}}\right)\vec{\nabla} G\right]\,\mathrm{d}A\nonumber\\ 
&+&\dfrac{1}{4\pi i k}\oint_{C}\vec{\nabla} G\left(\vec{\mathcal{H}}\cdot\mathrm{d}\vec{s}\right)
\end{eqnarray}
and
\begin{eqnarray}
\label{eq6}
\vec{H}(\vec{r})=
&\dfrac{1}{4\pi}&\int_{S}\left[ik\left(\vec{\mathcal{E}}\times\vec{n}\right)G
+\left(\vec{n}\times\vec{\mathcal{H}}\right)\times\vec{\nabla} G
+\left(\vec{n}\cdot{\vec{\mathcal{H}}}\right)\vec{\nabla} G\right]\,\mathrm{d}A\nonumber\\ 
&-&\dfrac{1}{4\pi i k}\oint_{C}\vec{\nabla} G\left(\vec{\mathcal{E}}\cdot\mathrm{d}\vec{s}\right),
\end{eqnarray}
where the Green function $G$ is given for example in \cite{Stratton1939, Török2000, Godana2023}. For both fields the first integral is the \textit{surface}, the second is the \textit{contour} term.

\begin{figure}[h]
\centering
\includegraphics[width=7.5 cm]
{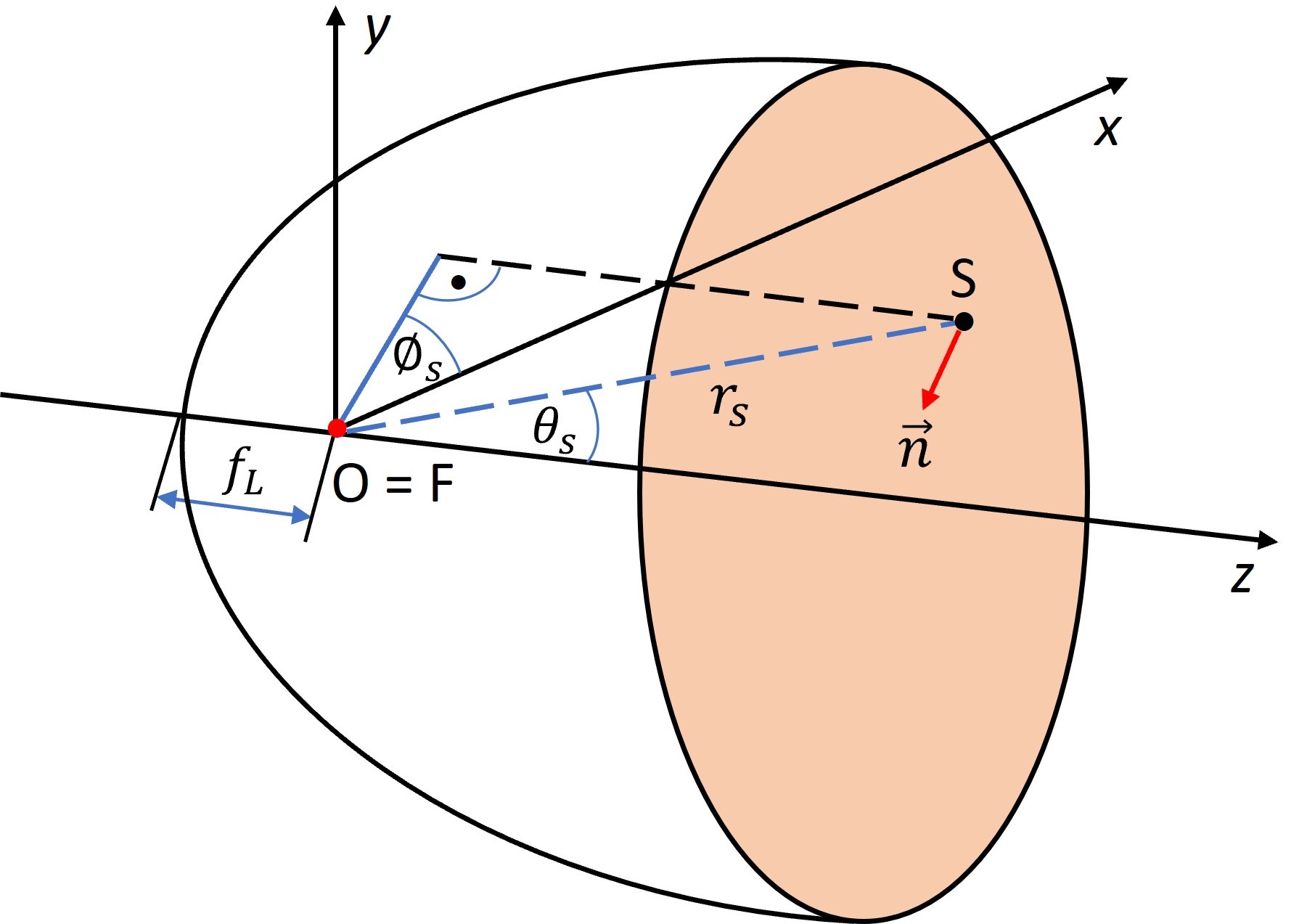}
\caption{Illustration of the parabolic mirror with notations.\label{fig2}}
\end{figure}   

We are curious about the complex electric and magnetic fields at an arbitrary point inside the parabolic mirror -- because of practical interest, especially in its focal region.
For the derivations, the electromagnetic boundary conditions concerning a perfectly reflecting surface, together with relations concerning the differential geometry of the paraboloid \cite{Török2000, Godana2023} were used. 

Supposing a Vectorial Gaussian incident beam as described by Eqs.\ (\ref{eq1}) and (\ref{eq2}) the electric and magnetic fields using Cartesian components, but cylindrical arguments are:
\begin{eqnarray}
\label{eq18}
\vec{\mathcal{E}}_i(\rho,z)&=&\left[a(\rho,z,z_w)\cos\phi,a(\rho,z,z_w)\sin\phi, b(\rho,z,z_w)\right]e^{-ikz}, \nonumber\\ 
\vec{\mathcal{H}}_i(\rho,z)&=&\left[a(\rho,z,z_w)\sin\phi,-a(\rho,z,z_w)\cos\phi, 0\right]e^{-ikz},
\end{eqnarray}
where in accordance with Eqs.\ (\ref{eq1}) and (\ref{eq2})
\begin{align}
\label{eq19}
a(\rho,z,z_w)=&\;\mathcal{E}_0 \frac{\rho}{w_0} [f(z-z_w)]^2 \exp\left(-f(z-z_w)\frac{{\rho}^2}{w_0^2}\right), \nonumber\\ 
b(\rho,z,z_w)=&\;i\theta_0 \mathcal{E}_0 [f(z-z_w)]^2\left(1-f(z-z_w)\frac{{\rho}^2}{w_0^2} \right)\cdot\exp\left(-f(z-z_w)\frac{{\rho}^2}{w_0^2}\right),
\end{align}
where the $f$ function is defined in Eq.\ (\ref{eq3}). The role of the $z_w$ parameter is to shift the beam waist from $z=0$ to $z=z_w$ (by replacing $z$ with $z-z_w$ in the argument of $f$) according to the circumstances in Subsection 4.2. 

Owing to the cylindrically symmetric illumination $a$ and $b$ along the surface $S$ can be regarded as only the function of the $\theta_s$ coordinate (see Fig.\ \ref{fig2}) and the $z_w$ parameter as:  
\begin{align}
\label{eq20}
a(\theta_s,z_w)=&\;\mathcal{E}_0 \frac{\rho_s(\theta_s)}{w_0} \left\{f\left[z_s(\theta_s)-z_w\right]\right\}^2\cdot\exp\left\{-f\left[z_s(\theta_s)-z_w\right]\left[\frac{\rho_s(\theta_s)}{w_0}\right]^2\right\}, \nonumber\\ 
b(\theta_s,z_w)=&\;i\theta_0 \mathcal{E}_0 \left\{f\left[z_s(\theta_s)-z_w\right]\right\}^2 \cdot\left\{1-f\left[z_s(\theta_s)-z_w\right]\left[\frac{\rho_s(\theta_s)}{w_0}\right]^2 \right\}\cdot \nonumber\\
&\cdot \exp\left\{-f\left[z_s(\theta_s)-z_w\right]\left[\frac{\rho_s(\theta_s)}{w_0}\right]^2\right\}.
\end{align}

After careful derivations, one can obtain
\begin{align}
\label{eq22}
\vec{E}_S(\rho,z)=&\frac{\exp(2ikf_L)}{2\pi}ik\int_{\delta}^{\pi}\int_{0}^{2\pi}\mathrm{d}\theta_s\,\mathrm{d}\phi_s\frac{\exp[ik(u-r_s)]}{u}r_s^2\sin\theta_s\cdot \nonumber\\
&\cdot
\left\{
\left[a\,\cos\phi_s+\left[ b-a\,\cot\left(\frac{\theta_s}{2}\right)\left(1-\frac{1}{iku}\right)\frac{\Delta x}{u}\right] \right]\vec{e}_{\rho}+
\right.
\nonumber\\
&+
\left.\left[a\,\cot\left(\frac{\theta_s}{2}\right)+\left[ b-a\,\cot\left(\frac{\theta_s}{2}\right)\left(1-\frac{1}{iku}\right)\frac{\Delta z}{u}\right] \right]\vec{e}_{z}\right\}
\end{align}
for the surface, and
\begin{align}
\label{eq24}
\vec{E}_C(\rho,z)=&-\frac{f_L\exp(2ikf_L)}{\pi}\cot\left(\frac{\delta}{2}\right)\oint a\frac{\exp[ik(u-r_s)]}{u}\left(1-\frac{1}{iku}\right)\cdot \nonumber\\
&\cdot\left(\frac{\Delta x}{u}\vec{e}_{\rho}+\frac{\Delta z}{u}\vec{e}_{z}\right)\,\mathrm{d}\phi_s
\end{align}
for the contour electric field term, where $\Delta x=r_s\sin\theta_s\cos\phi_s-\rho$, $\Delta y=r_s\sin\theta_s\sin\phi_s$, $\Delta z=r_s-2f_L-z$, $u=\left(\Delta x^2+\Delta y^2+\Delta z^2 \right)^{\frac{1}{2}}$ and $\vec{e}_{\rho}$, $\vec{e}_z$ are the unit vectors.

At the focus, ($\rho=z=0$) a fully analytical expression can be obtained for the contour term:
\begin{equation}
\label{eq25}
\vec{E}_C(\mathrm{F})=-\frac{a(\delta,z_w)}{2}\exp(2ikf_L)\sin(2\delta)\left(1-\frac{1-\cos\delta}{2ikf_L}\right)\vec{e}_z.
\end{equation}
This expression is especially useful if we are curious about the electric field at the focus in that case when the contour term has dominance. 

Using Eqs.\ (\ref{eq22}) and (\ref{eq24}) the total complex electric field can be given as:
\begin{equation}
\label{eq26}
\vec{E}(\rho,z)=\vec{E}_S(\rho,z)+\vec{E}_C(\rho,z).
\end{equation}

For the magnetic field having solely azimuthal component, one obtains:
\begin{align}
\label{eq29}
\vec{H}(\rho,z)=&\frac{\exp(2ikf_L)}{2\pi}ik\int_{\delta}^{\pi}\int_{0}^{2\pi}\mathrm{d}\theta_s\,\mathrm{d}\phi_s\, a \frac{\exp[ik(u-r_s)]}{u}r_s^2\sin\theta_s\cdot \nonumber\\
&\cdot\left(1-\frac{1}{iku}\right)\left[\cot\left(\frac{\theta_s}{2}\right)\frac{\Delta x}{u}-\cos\phi_s\frac{\Delta z}{u}\right]\vec{e}_{\phi},
\end{align}
where apart of the $\vec{e}_{\phi}$ azimuthal unit vector all symbols are explained above.

In the next Section, the electric and magnetic fields will be analyzed in the focal region based on the formulae derived in this Section.

\section{Analyses of the Electromagnetic Field in the Focal Region}
\noindent

In the following, as a practical choice, the parabolic mirror will be considered as a segment of paraboloid in the $\frac{\pi}{3}<\theta_s<\pi$ region (see Fig.\ \ref{Fig3ab}). For the sake of generality, instead of specifying the focal length $f_L$, the wavelength $\lambda$ and $w_0$ as absolute parameters, we use the $\lambda/f_L$ and $w_0/f_L$ relative parameters. We note, that the effect of the $\lambda/f_L$ ratio on the electric field distributions has been already discussed in detail \cite{Godana2023}. During the following analysis $\lambda/f_L<0.1$ is assumed since the $\lambda/f_L>0.1$ range is uninteresting from a practical point of view.

\subsection{Incident Beam with negligible Divergence}
\noindent

In many practical cases, the divergence of the beam incident on the focusing element is negligible. This means the limit of $\theta_0 \rightarrow 0$ in Eq.\ (\ref{eq1}), in other words, the Rayleigh range is much larger than the typical size of the parabolic mirror i.\ e.\ $\frac{z}{z_0} \rightarrow 0$ in the formulae. Accordingly, we can use the $f(z) \approx 1$, and $w(z) \approx w_0$ approximations leading to
\begin{eqnarray} \label{eq30}
\begin{split}
 &a(\rho ,z)=a(\rho)=\mathcal{E}_0 \frac{\rho}{w_0} \exp\left(-\frac{{\rho}^2}{w_0^2}\right),\\
&b(\rho ,z)=0.   
\end{split}
\end{eqnarray}

\subsubsection{The Fields at the Focus}
\noindent

\vspace{5mm}

\begin{figure} [h]
\centering
\includegraphics[width=12.5 cm]
{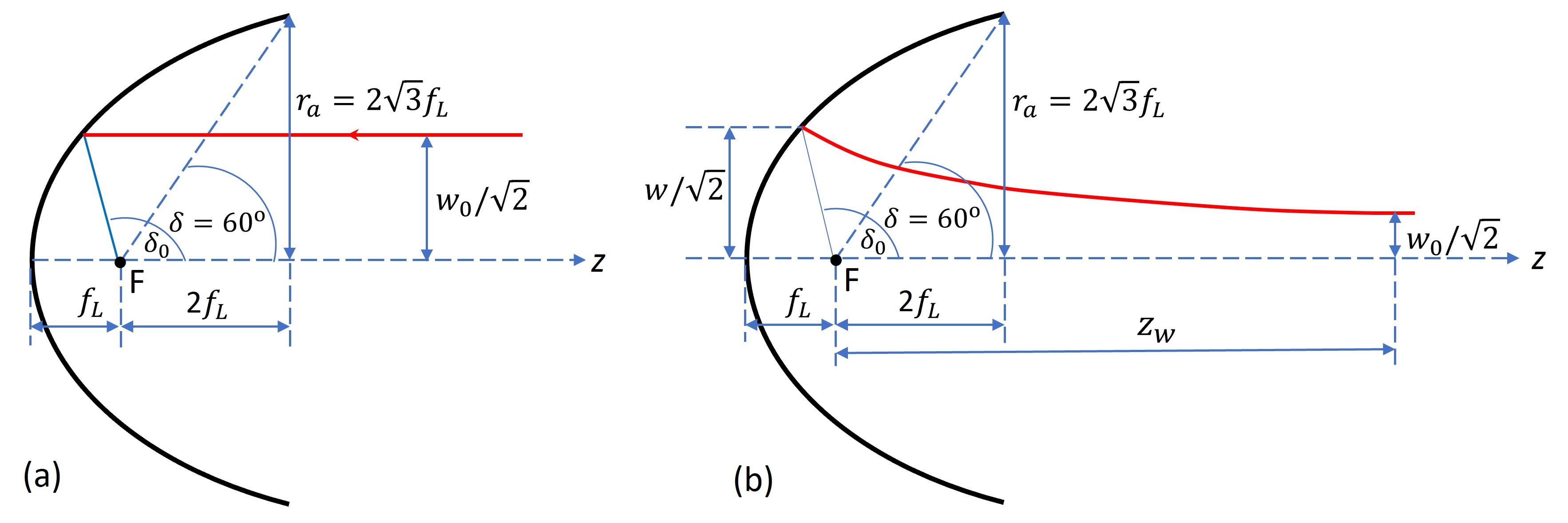}
\caption{The geometry of the incidence on the parabolic mirror for a beam with negligible (a), and nonnegligible divergence (b).\label{Fig3ab}}
\end{figure}

Let us assume, that a Gaussian vector beam with negligible divergence given by Eqs.\ (\ref{eq1}), (\ref{eq2}) and (\ref{eq30}) is incident on the parabolic mirror as can be seen in Fig.\ \ref{Fig3ab}a. $w_0$ relates to the $\delta_0$ focusing angle (see Fig.\ \ref{Fig3ab}a) as
\begin{equation}
\label{eq31}
\frac{w_0}{\sqrt{2}}=\frac{2f_L}{1-\cos\delta_0}\sin\delta_0.
\end{equation}

\begin{figure}
\centering
{\includegraphics[width=10.5 cm]{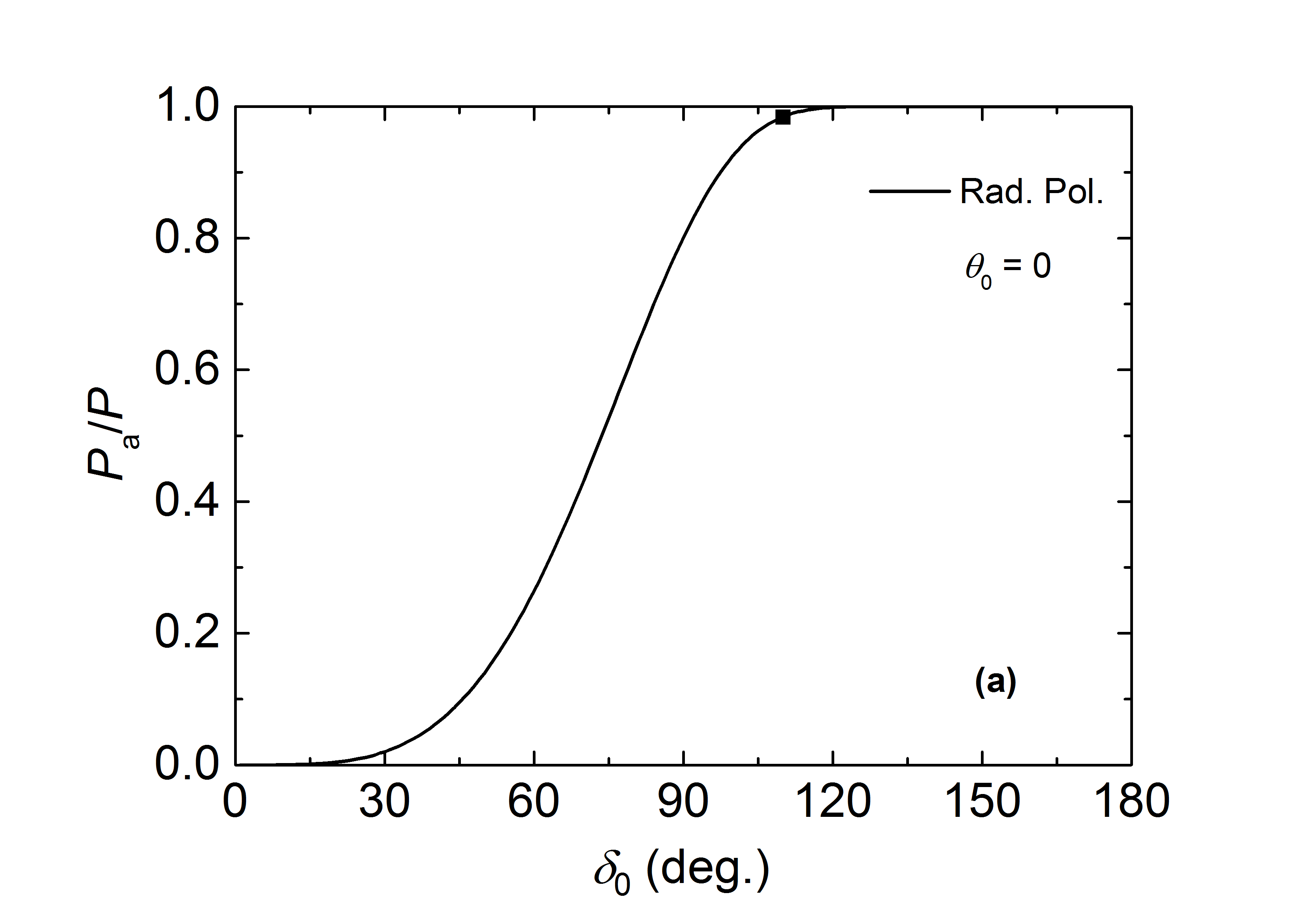}}
{\includegraphics[width=10.5 cm]{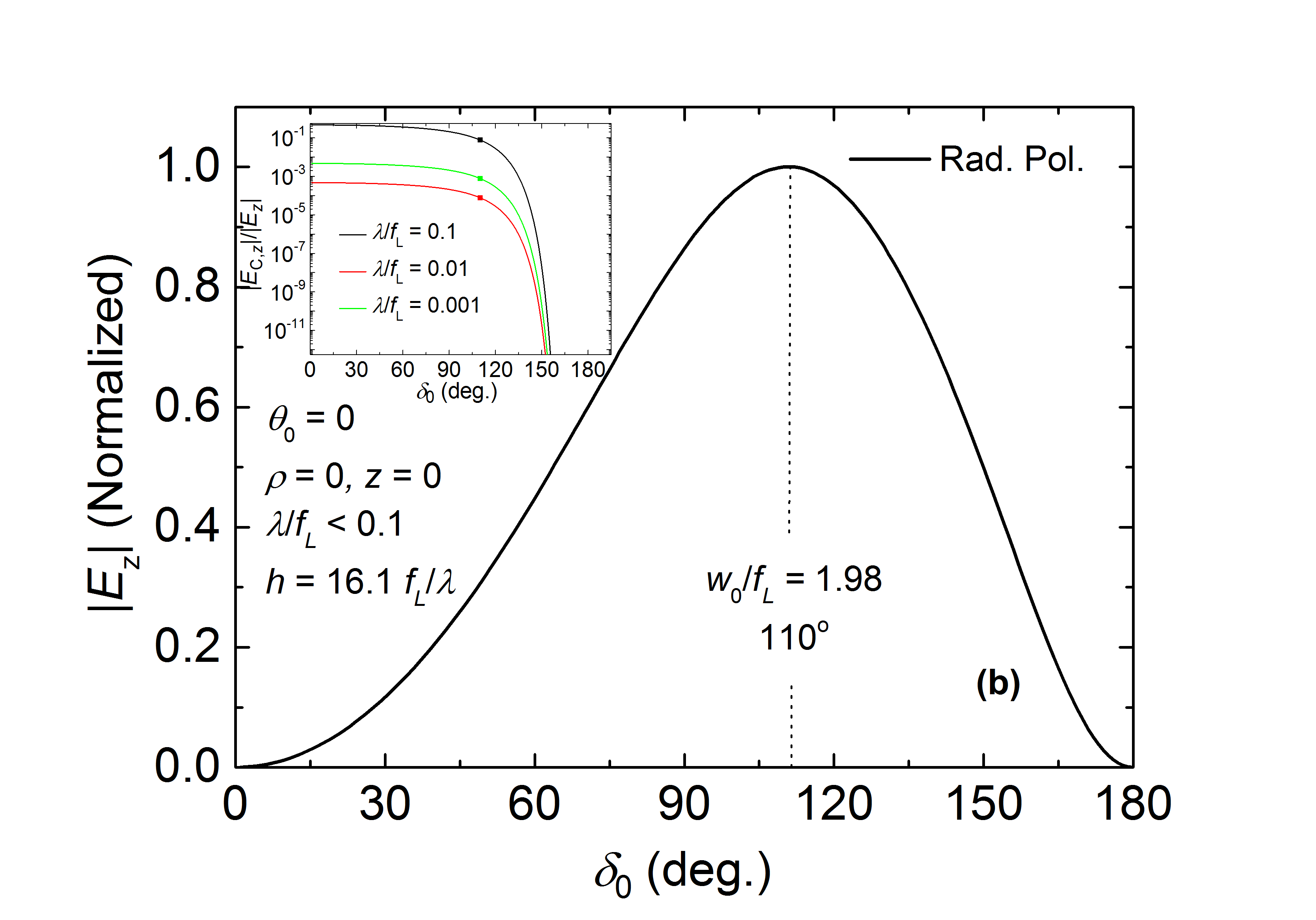}}
{\includegraphics[width=10.5 cm]{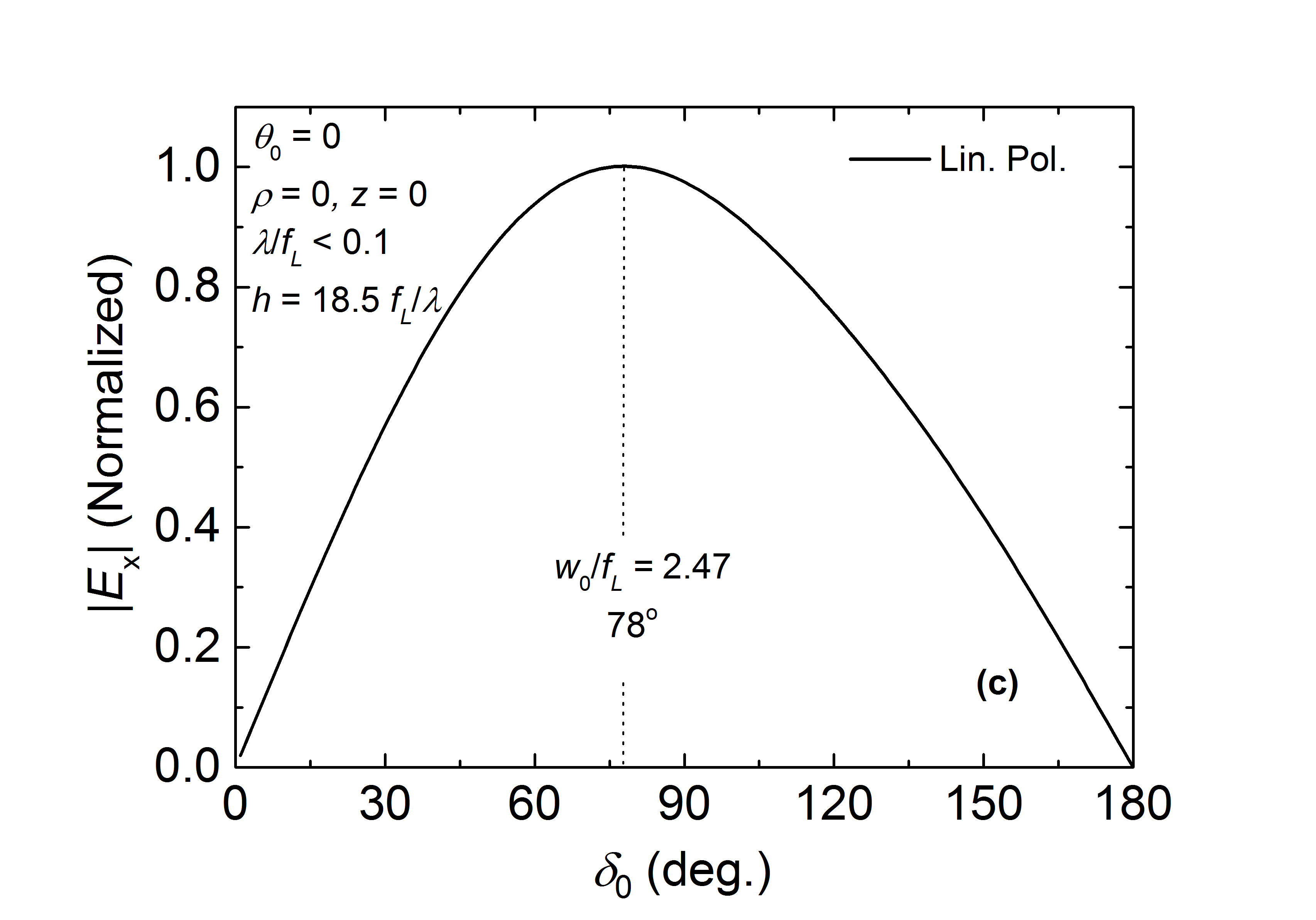}}
\caption{The beam power ratio falling into the aperture of the parabolic mirror (a). The longitudinal (b) and transversal (c) electric field amplitudes at the focus for radially (b) and linearly (c) polarized Gaussian beam versus the $\delta_0$ focusing angle. The inset (b) shows the amplitude ratio of the contour term to the total field for $\lambda/f_L=0.1,\,0.01$ and $0.001$. We draw attention to the logarithmic scale.\label{fig4a&b}}
\end{figure}

The $P_a$ beam power selected by the parabolic mirror with its $r_a=r_s(\theta_s=60^{\circ})=2\sqrt{3}f_L$ aperture radius relative to the total beam power $P$ is depicted in Fig.\ \ref{fig4a&b}a versus the $\delta_0$ angle.  
  
 At first, we concentrate on the fields exactly at the focus. The magnetic field is zero, and the electric field has only a $z$ component. We determined $\vert E_z \vert$ at the focus ($\rho=z=0$) from Eqs.\ (\ref{eq22}), (\ref{eq25}),  (\ref{eq26}) and (\ref{eq30}). According to the special selection of the paraboloid segment, the lower bound of the integration (concerning $\theta_s$ in Eq.\ (\ref{eq22})) was $\delta=\frac{\pi}{3}$. In Fig.\ \ref{fig4a&b}b $\vert E_z \vert$ is plotted versus the $\delta_0$ parameter (related to the beam radius according to Eq.\ (\ref{eq31})). The peak is normalized to 1, thereby obtaining a curve to be practically independent of $\lambda/f_L$ (if $\lambda/f_L<0.1$) \cite{Godana2023}.

The peak belongs to $\delta_0=110^{\circ}$ (see Fig.\ \ref{Fig3ab}a) with corresponding value of $w_0/f_L = 1.98$. In this case, $98.4\%$ of the total beam power is transmitted through the aperture of the parabolic mirror (see the black square in Fig.\ \ref{fig4a&b}a). From a practical point of view, it is important to know, how the focused field amplitude relates to the total beam power. Therefore, we introduced an average electric field $E_{f}$, which scales with the total beam power. $E_{f}$ is fixed by the condition to get power equivalence between the Gaussian beam going to be focused and a fictive Flat-top beam with uniform amplitude $E_{f}$ and beam radius $r_a$ \cite{Pálfalvi2023}, namely
\begin{equation}
\label{Eaverage}
\frac{1}{2}c\epsilon_0 E_{f}^2 r_a^2 \pi=6\pi c\epsilon_0 E_{f}^2 f_L^2 = P \left(=\frac{1}{8}c\epsilon_0 \mathcal{E}_{0}^2 w_0^2 \pi\right) .
\end{equation}

The field enhancement factor relative to this fictive average field (used for characterization in the following) is $h=\vert E_z\vert/E_{f}$. For the peak point of Fig.\ \ref{Fig3ab}a $h=16.1\cdot f_L/\lambda$ (if $\lambda/f_L<0.1$) was found. This enhancement value is only $\sim$10$\%$ lower than the $h_{\mbox{max}}=E_{\mbox{max}}/E_f=\sqrt{32\pi}\cdot f_L/\lambda=17.8\cdot f_L/\lambda$ theoretical maximum obtained for ideal dipole wave in $4\pi$ focusing geometry \cite{Bassett1986}. The shorter wavelength is preferable for a given $f_L$.

As an example supposing $f_L=50\, \mathrm{mm}$ (a typical value) the assumed $\lambda/f_L<0.1$ condition holds not only for the visible and (near-, mid-) infrared but also for the THz frequency range (0.1 - 10 THz). For $\lambda/f_L=0.1$ the corresponding frequency is only 0.06 THz, so if $\lambda/f_L<0.1$ the whole THz range is covered. The importance of the THz fields is outstanding due to their applicability for particle acceleration  \cite{Zhang2018,Nanni2015,Hibberd2020} because of their advantageous wavelength and because of the availability of pulses with extremely high pulse energies and electric field strengths owing to the tilted-pulse-front pumping technique \cite{Zhang2021,Fülöp2014}. For example at 0.6 THz frequency (typically available with $\mathrm{LiNbO}_3$ nonlinear crystals) an enhancement factor as large as $h=1610$ can be reached. This is close to the $h$ value estimated by less sophisticated approximations under very similar geometrical circumstances \cite{Pálfalvi2023}. Supposing for example an incident THz electric field of $100$ kV/cm, a longitudinal electric field of $161$ MV/cm becomes available at the focus, which is excellent for particle acceleration applications.  

In many cases, depending on the focusing geometry and the wavelength, the contribution of the contour integral term (Eq.\ (\ref{eq24})) is negligible compared to the surface term (Eq.\ (\ref{eq22})). In Fig.\ \ref{fig4a&b}b in the inset one can see the $\vert E_{C,z}\vert / \vert E_{z}\vert$ amplitude ratio of the contour term to the total axial field versus $\delta_0$ for $\lambda / f_L=0.1,\,0.01$ and $0.001$. It is seen, that this ratio decreases with increasing $\delta_0$. This behavior is understandable since the radius of the high-intensity part of the illuminated paraboloid decreases with $\delta_0$ and consequently, the field decreases along the contour. The particular case belonging to the peak (at $110^{\circ}$) of the main curve of Fig.\ \ref{fig4a&b}b is shown by a square.

It is also informative and useful to make a comparison with the focusing of the linearly polarized Gaussian beam. Therefore, we determined $\vert E_x \vert$ at the focus ($\rho=z=0$). In Fig.\ \ref{fig4a&b}c the normalized $\vert E_x \vert$ is plotted versus $\delta_0$. Note, that in this case the relationship between $w_0$ and $\delta_0$ corresponding to Eq.\ (\ref{eq31}) reads as:
\begin{equation}
\label{eq32}
w_0=\frac{2f_L}{1-\cos\delta_0}\sin\delta_0,
\end{equation}
since for a linearly polarized beam $w_0$ is regarded as the characteristic size instead of $w_0/\sqrt{2}$ as mentioned above.
The peak of the curve is reached at $\delta_0=78^{\circ}$, with corresponding beam waist of $w_0/f_L = 2.47$. In this case, $98\%$ of the total beam power is transmitted through the aperture of the parabolic mirror (which is very close to the above-mentioned value of $98.8\%$ for the case of a radially polarized beam), hence providing a basis for making the comparison with the radially polarized beam. The field enhancement factor is $h=18.5\cdot f_L/\lambda$. This value just exceeds the corresponding $h$ for the radial polarization also underlining the effectiveness of generating a longitudinal electric field.

\begin{figure}[h]
\centering
\includegraphics[width=10.5 cm]
{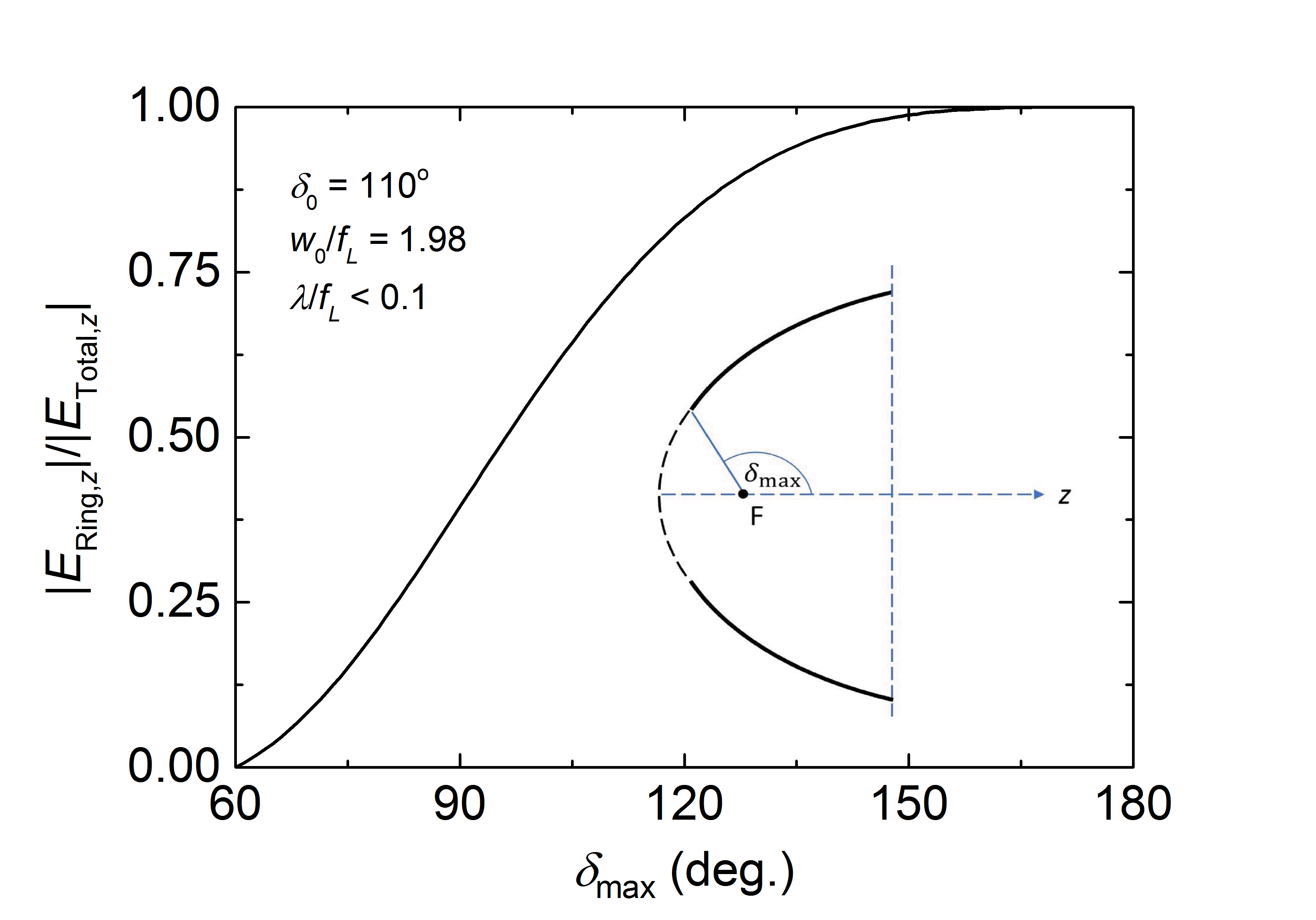}
\caption{The amplitude of the longitudinal field component concerning the ring relative to the field amplitude achievable with the entire paraboloid (at the focus) versus $\delta_{\mathrm{max}}$ (see the inset). The left edge of the curve corresponds to the thin ring, while its right edge to the continuous paraboloid.\label{fig5}}
\end{figure}

Let us turn back to the radial polarization. When the peak point (belonging to the position of $\delta_0=110^{\circ}$ in Fig.\ \ref{fig4a&b}b) was calculated, the integration range in $\theta_s$ was encompassed from $60^{\circ}$ to $180^{\circ}$. In a thought experiment considering, for example, the advantageous illumination geometry belonging to $\delta_0=110^{\circ}$ (with corresponding $w_0/f_L=1.98$) let us keep the lower bound of the integration range at a value of $60^{\circ}$, but vary the upper bound, $\delta_{\mathrm{max}}$ between $60^{\circ}$ and $180^{\circ}$. Thereby a ring-like paraboloid segment (as illustrated in the inset of Fig.\ \ref{fig5}) is obtained bounded by two contours (at $60^{\circ}$ and $\delta_{\mathrm{max}}$, respectively). In Fig.\ \ref{fig5} the amplitude of the longitudinal electric field at the focus concerning the ring relative to the field amplitude achievable with the entire ($60^{\circ}<\theta_s<180^{\circ}$) paraboloid is plotted versus $\delta_{\mathrm{max}}$. During the calculations, this ring-like structure was taken into consideration by the difference of two contour terms  (Eq.\ (\ref{eq24})). It is obvious from the graph, that -- in this particular focusing geometry -- the region of the paraboloid between 150 and $180^{\circ}$ has practically no contribution to the total electric field. This result is important information for specialists dealing with the development of particle accelerators. The ‘hole’ on the paraboloid around its vertex does not detract from the available longitudinal electric field, and might have a practical purpose as well: it ensures the unobstructed particle transfer. The details of the ring-like paraboloid setup are discussed in Ref.\ \cite{Pálfalvi2023}.

\subsubsection{The Fields at the Vicinity of the Focus}
\noindent

\vspace{5mm}

\begin{figure}[h]
\centering
\includegraphics[width=10.5 cm]{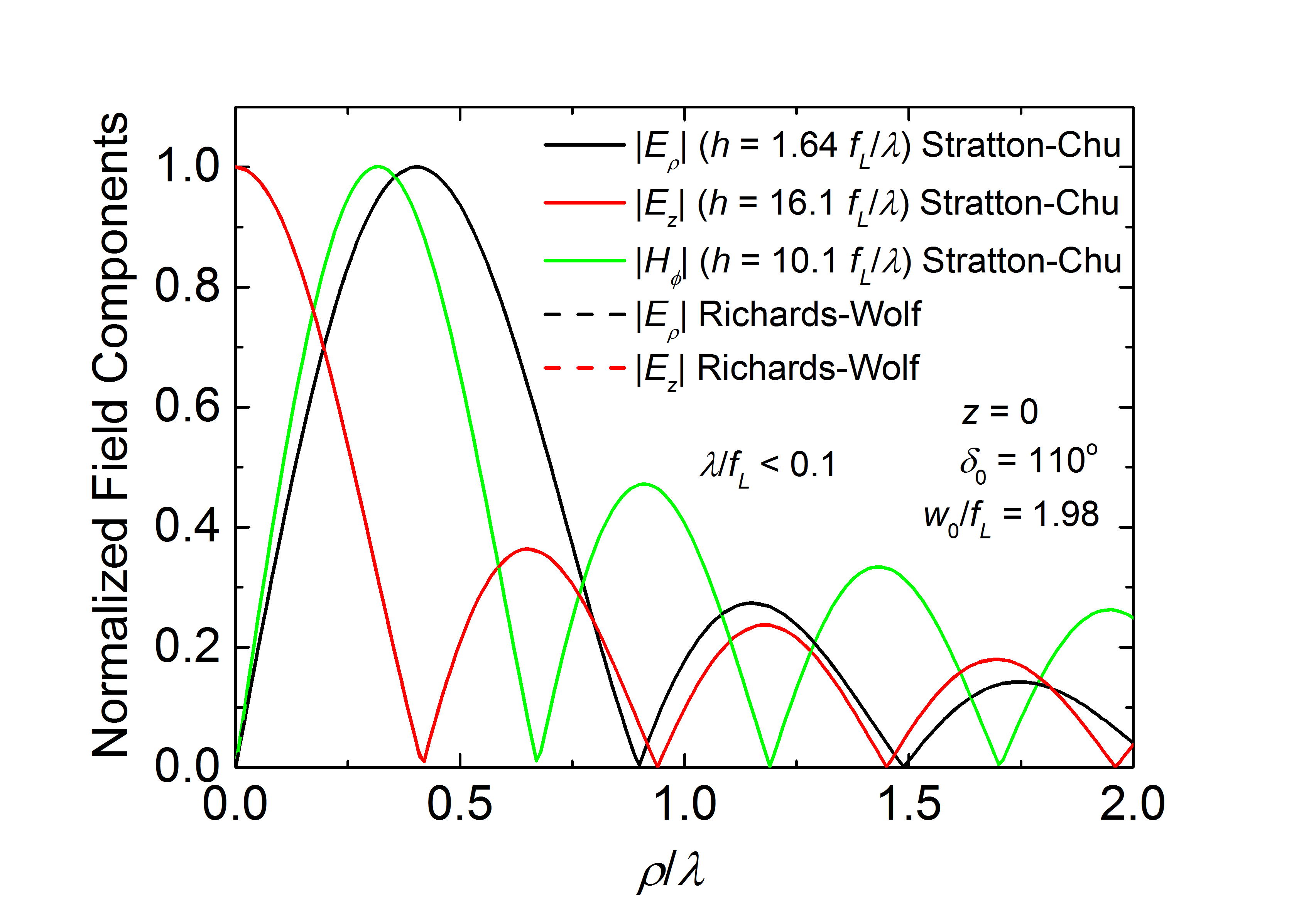}
\caption{The radial electric (black line), the axial electric (red line), and the azimuthal magnetic (green line) field components versus the radial distance from the focus for $z=0$. The horizontal scale is normalized by the wavelength, and the peaks of the curves are normalized to unity. The field enhancement factors, $h$ are indicated in the graph. Note that the dashed curves used for validation fully coincide with the corresponding solid curves. \label{fig6}}
\end{figure}

It was examined, how the amplitude of the field components vary with the distance from the focus in the radial direction. The behavior of the radial (black line) and axial (red line) electric field and the azimuthal magnetic field (green line) are plotted in Fig.\ \ref{fig6}. The horizontal scale is normalized by the wavelengths, and the peaks of all the curves are normalized to 1. Advantageously, by such normalization, the curves do not change with the $\lambda/f_L$ ratio in the $\lambda/f_L<0.1$ range. The illumination parameters were again $\delta_0=110^{\circ}$, $w_0/f_L=1.98$.

As can be seen, all field components oscillate with a decaying amplitude. The spatial oscillation period is somewhat shorter for $\vert H_{\phi} \vert$ than for $\vert E_{\rho} \vert$ leading to a continuous slight phase shift with the $\rho$ distance. The position of the axial electric field amplitude maxima and the magnetic field minima (and vice versa) approximately coincide. Except for the first (for the axial electric field the first half) period, the period of the spatial oscillation approximately equals $\lambda/2$ as it is expected for a standing wave. Considering the enhancement factors, for the radial electric field component, $\vert E_{\rho} \vert$ it is the lowest ($h=1.64\cdot f_L/\lambda$), and for the axial electric field, $\vert E_{z} \vert$ it is the largest ($h=16.1\cdot f_L/\lambda$). For a typical value of $f_L/\lambda=100$ (as also supposed in the example above) these enhancement factors are 164, 1610, and 1010 for $\vert E_{\rho} \vert$, $\vert E_{z} \vert$ and $\vert H_{\phi} \vert$, respectively. The $\int_0^{\infty}\left(\vert E_{\rho} \vert^2+\vert E_{z} \vert^2\right)\rho\,\mathrm{d}\rho=\int_0^{\infty}\vert H_{\phi} \vert^2\rho\,\mathrm{d}\rho$ relation was verified numerically.

\begin{figure}[h]
\centering
\includegraphics[width=11 cm]
{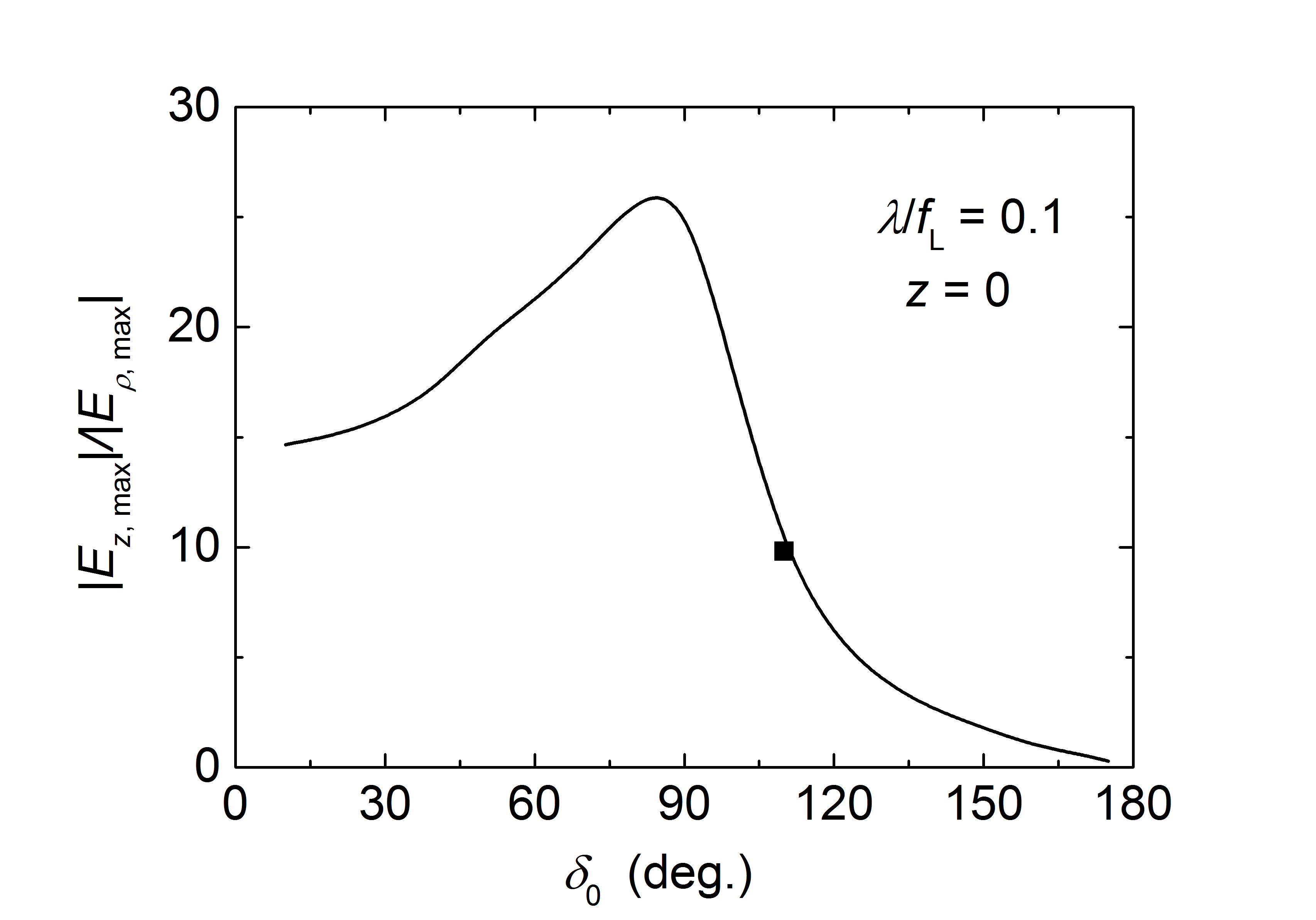}
\caption{The amplitude ratio of the maximal axial to the maximal radial electric field component versus the focusing angle. \label{fig7}}
\end{figure}

The amplitude ratio of the maximal axial, $\vert E_{z, \mathrm{max}}\vert$ to the maximal radial, $\vert E_{\rho, \mathrm{max}}\vert$ electric field component \cite{Youngworth2000} is plotted versus the $\delta_0$ focusing angle in Fig.\ \ref{fig7}. The curve belongs to the focal plane ($z=0$). Note, that $\vert E_{z, \mathrm{max}}\vert$ is reached on the axis ($\rho=0$), while $\vert E_{\rho, \mathrm{max}}\vert$ is reached at an off-axis ($\rho \ne 0$) point for any $\delta_0$ (similarly, as it is seen in Fig.\ \ref{fig6} for $\delta_0=110^{\circ}$). As it is seen, the $\vert E_{z, \mathrm{max}}\vert/\vert E_{\rho, \mathrm{max}}\vert$ ratio monotonically decreases with $\delta_0$. The case belonging to Fig.\ \ref{fig6} is indicated by a square symbol. The corresponding amplitude ratio is $h_z/h_{\rho}=16.1\cdot(f_L/\lambda)/1.64\cdot(f_L/\lambda)=9.81$. 

\begin{figure}[h]
\centering
\includegraphics[width=11 cm]
{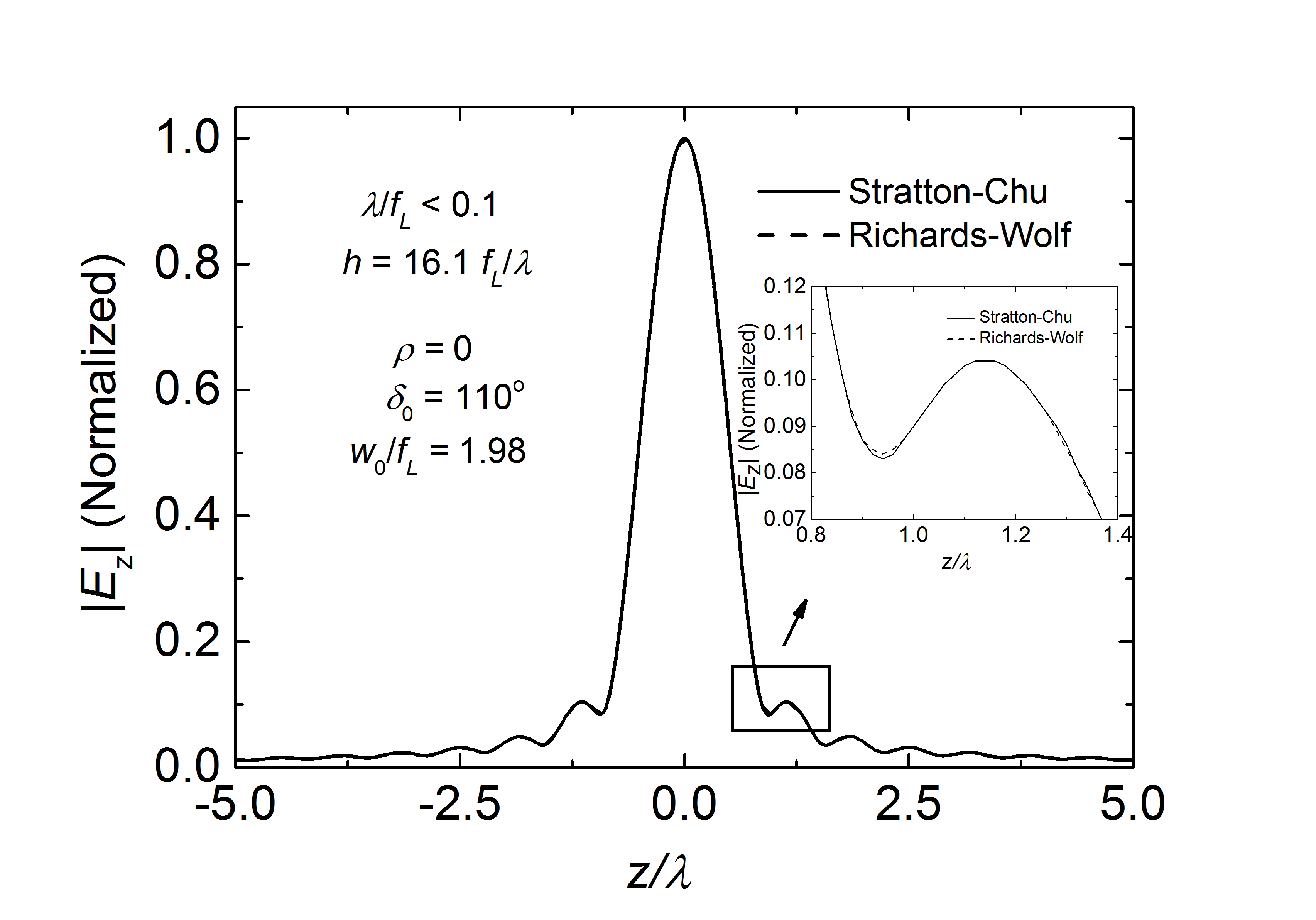}
\caption{The amplitude of the longitudinal electric field component versus the longitudinal coordinate, $z$ for $\rho=0$.  The horizontal scale is normalized by the wavelength, and the peaks of the curves are normalized to unity. The inset shows a magnified region of the main graph. The solid lines refer to our Stratton–Chu-based, while the dashed ones refer to the Richards-Wolf-based theory. Note that in the main part of the figure, the two types of lines fully coincide. \label{fig8}}
\end{figure}

It was also examined, how the fields vary in the $z$ direction. For simplicity, this was examined only for the longitudinal electric field, for $\rho=0$. $\vert E_z \vert$ normalized to 1 is plotted in Fig.\ \ref{fig8}. The horizontal scale is normalized by the wavelengths because of the reason mentioned above. The supposed focusing parameters are the same as for  Fig.\ \ref{fig6}. The curve is symmetric to $z=0$. It shows a quasi-oscillation nature with a spatial period more than twice that in the $\rho$ direction for the given $\delta_0$ (Fig.\ \ref{fig6}). For smaller $\delta_0$ this ratio is larger.

\begin{figure} [h]
\centering
\includegraphics[width=10.5 cm]
{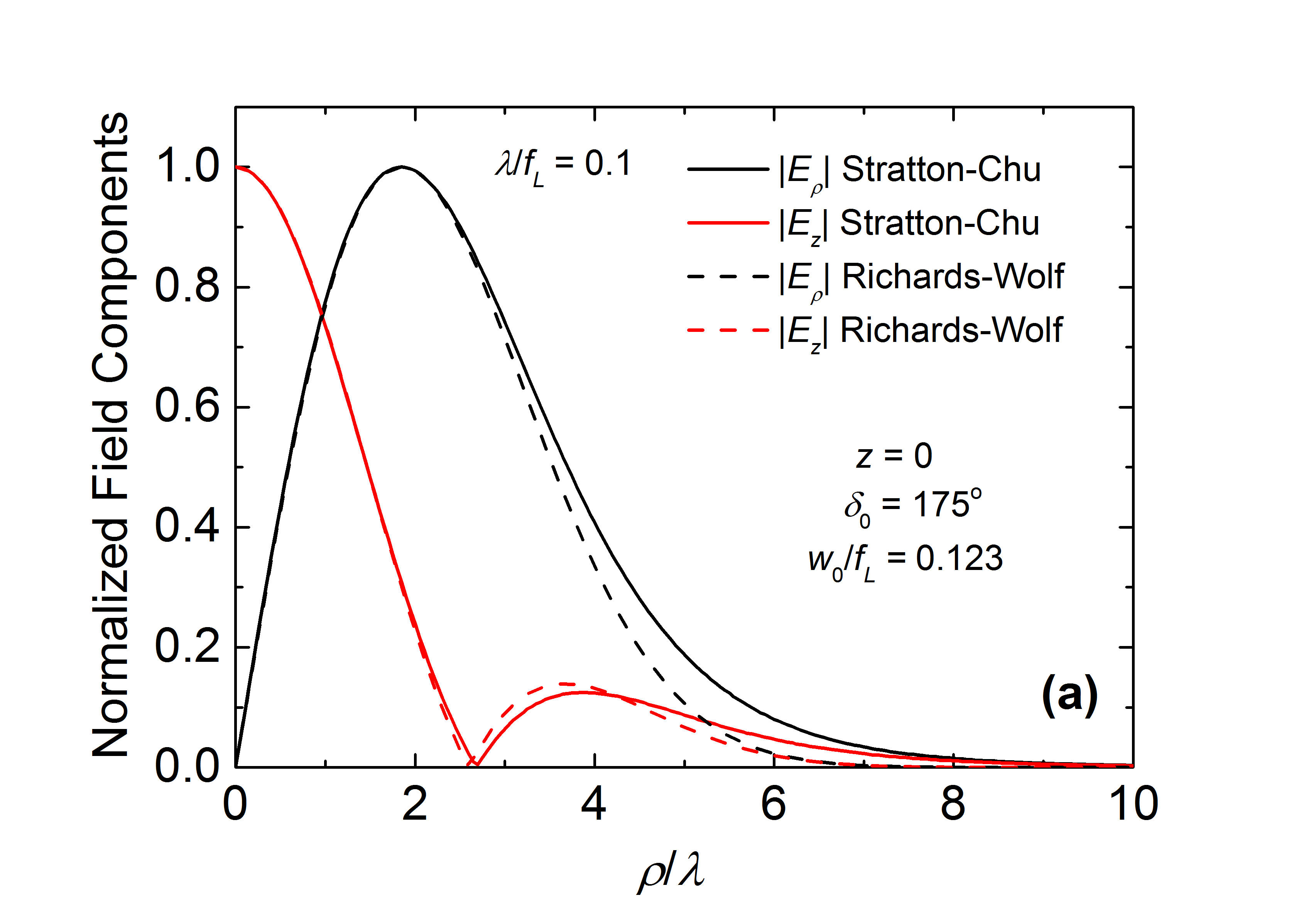}
\includegraphics[width=10.5 cm]
{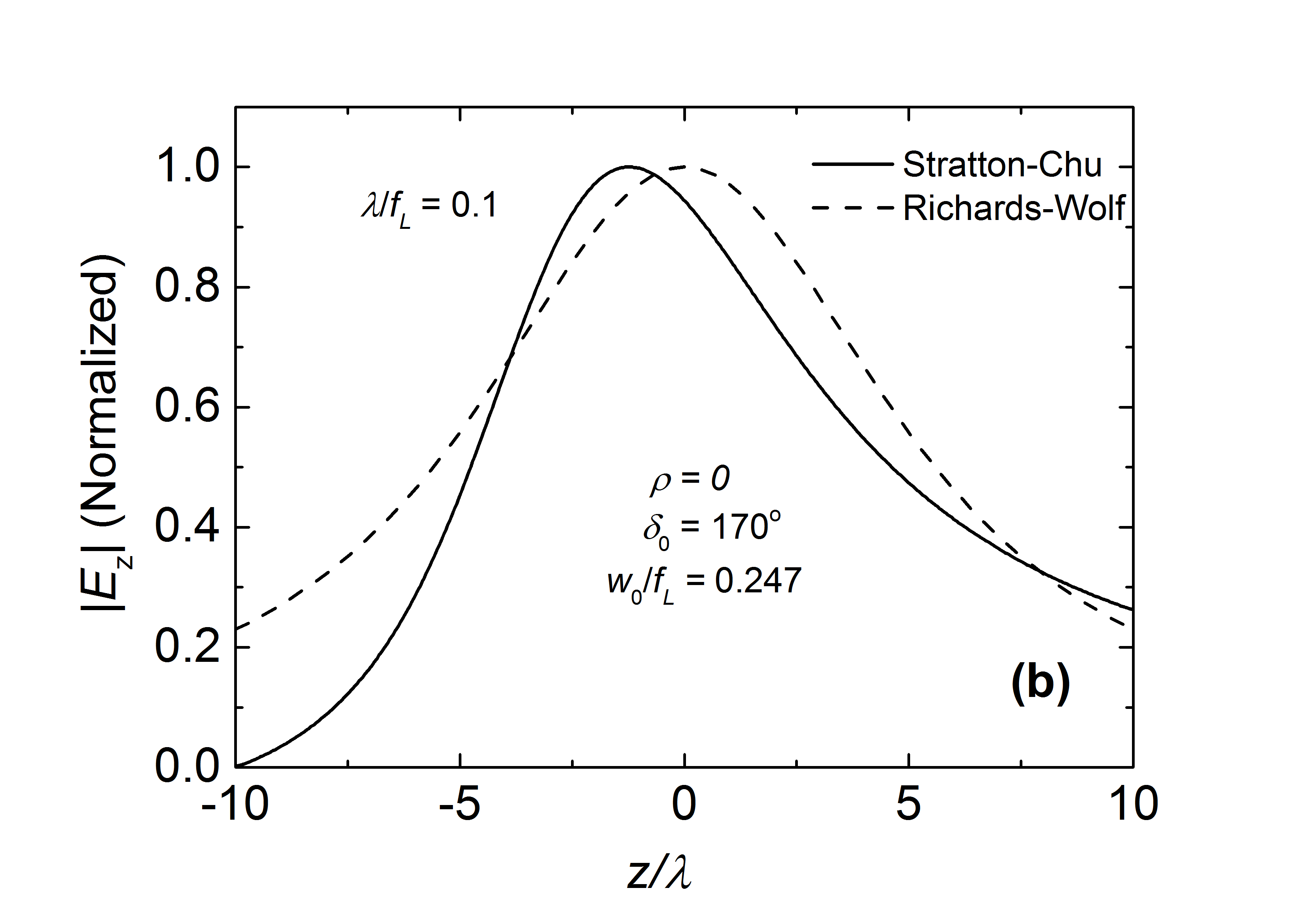}
\caption{The normalized transversal (a) and longitudinal (b) distribution of the electric field components around the focus for $\lambda/f_L=0.1$. The solid curves were computed by the Stratton-Chu-based, and the dashed ones by the Richards-Wolf-based theory. \label{fig9a&b}}
\end{figure}

We have compared our results calculated by the electric field formulae we derived from the Stratton–Chu theory with a Richards-Wolf theory \cite{Richards} based method \cite{Dehez2012}. The curves computed by using Eqs.\ 4a and b of \cite{Dehez2012} were added to Figs.\ \ref{fig6} and \ref{fig8}, respectively (dashed lines). In both figures, the dashed (Richards–Wolf) curves perfectly coincide with the corresponding solid (Stratton–Chu) curves. Their difference can be made visible only by magnification (as an example see the inset in Fig.\ \ref{fig8}) convincingly demonstrating the validity of our derivations. We have generally found, that the results obtained by the different concepts (Stratton–Chu/Richards–Wolf) are in agreement if the size of the high-intensity illuminated area is significantly larger than the wavelength. However, if the size of the high-intensity illuminated area is comparable to the wavelength a deviation appears between the curves obtained by the different theoretical concepts. For $\lambda/f_L=0.1$ the deviation in the radial field distribution is well observable for $\delta_0 \gtrsim 175^{\circ}$ (Fig.\ \ref{fig9a&b}a), while in the longitudinal field distribution the deviation appears even for $\delta_0 \gtrsim 170^{\circ}$ (Fig.\ \ref{fig9a&b}b). Therefore, in such cases, the use of the formalism developed by us is unavoidable.     

Certainly, the information provided by Figs.\ \ref{fig6}, \ref{fig7}  and \ref{fig8} on the field characteristics, and the expansion of the high field region around the focus is interesting for specialists designing particle acceleration by tightly focused fields.

\subsection{The Effect of the Beam Divergence}
\noindent

The divergence of the incident beam, $\theta_0$ which was neglected in the previous investigations is considered to be a finite parameter in the following. At a given $\lambda$ wavelength the $\theta_0$ parameter determines the beam waist and the Rayleigh range in the following way:
\begin{equation}
\label{eq33}
w_0(\theta_0)=\frac{\lambda}{\pi \theta_0}\,\,\,\, \mbox{and}\,\,\,\, z_0(\theta_0)=\frac{\lambda}{\pi \theta_0^2}.
\end{equation}
Let us suppose, that the position of the beam waist of the incident beam with a given $\theta_0$ divergence angle is at $z=z_w$ (Fig.\ \ref{Fig3ab}b). The relation between $w$ and the $\delta_0$ focusing angle (belonging to the $w/\sqrt{2}$ point on the paraboloid, as can be seen in Fig.\ \ref{Fig3ab}b) is:
\begin{equation}
\label{eq34}
\frac{w}{\sqrt{2}}=\frac{2f_L}{1-\cos\delta_0}\sin\delta_0.
\end{equation}
Hence, $w$ as the function of $\delta_0$ is:
\begin{equation}
\label{eq35}
w(\delta_0)=\frac{2\sqrt{2}f_L}{1-\cos\delta_0}\sin\delta_0.
\end{equation}
Using the relation
\begin{equation}
\label{eq36}
w(\delta_0)=w_0\sqrt{1+\left(\frac{z_w-r_s(\delta_0)\cos\delta_0}{z_0(\theta_0)}\right)^2}
\end{equation}
one obtains
\begin{equation}
\label{eq37}
z_w(\delta_0,\theta_0)=\frac{2f_L}{1-\cos\delta_0}\cos\delta_0+z_0(\theta_0)\sqrt{\left(\frac{w(\delta_0)}{w_0(\theta_0)}\right)^2-1}.
\end{equation}
So, for given $f_L$ and given  $\theta_0$ and $\delta_0$ parameters the incident beam becomes known according to Eqs.\ (\ref{eq33}) and (\ref{eq37}).  

\begin{figure} [h]
\centering
\includegraphics[width=10.5 cm]
{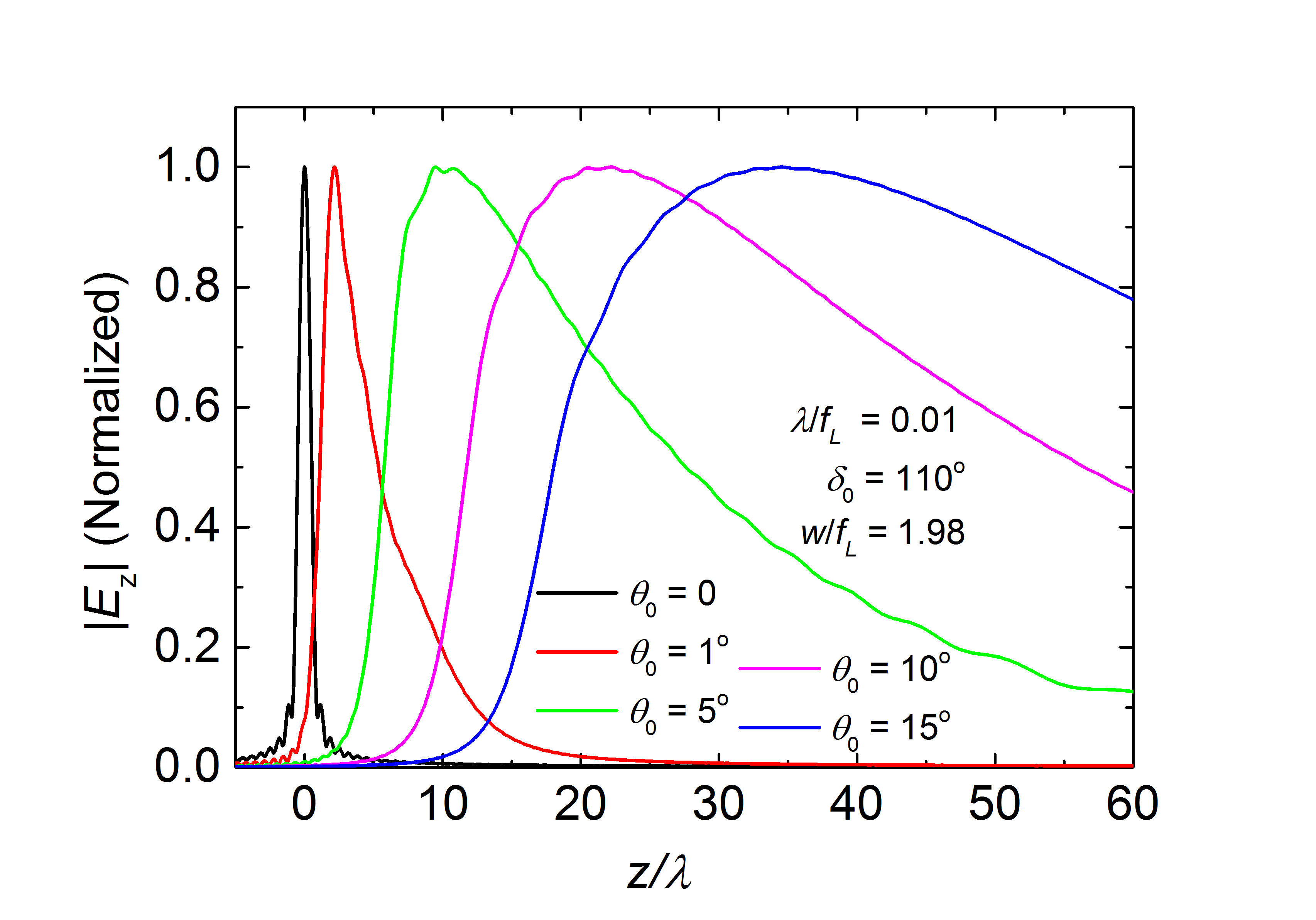}
\caption{The amplitude of the longitudinal electric field component versus the longitudinal coordinate, $z$ for $\rho=0$, and different $\theta_0$ values. The horizontal scale is normalized by the wavelength, the peaks are normalized to unity. \label{fig10}}
\end{figure}

\begin{figure} [h]
\centering
\includegraphics[width=10.5 cm]
{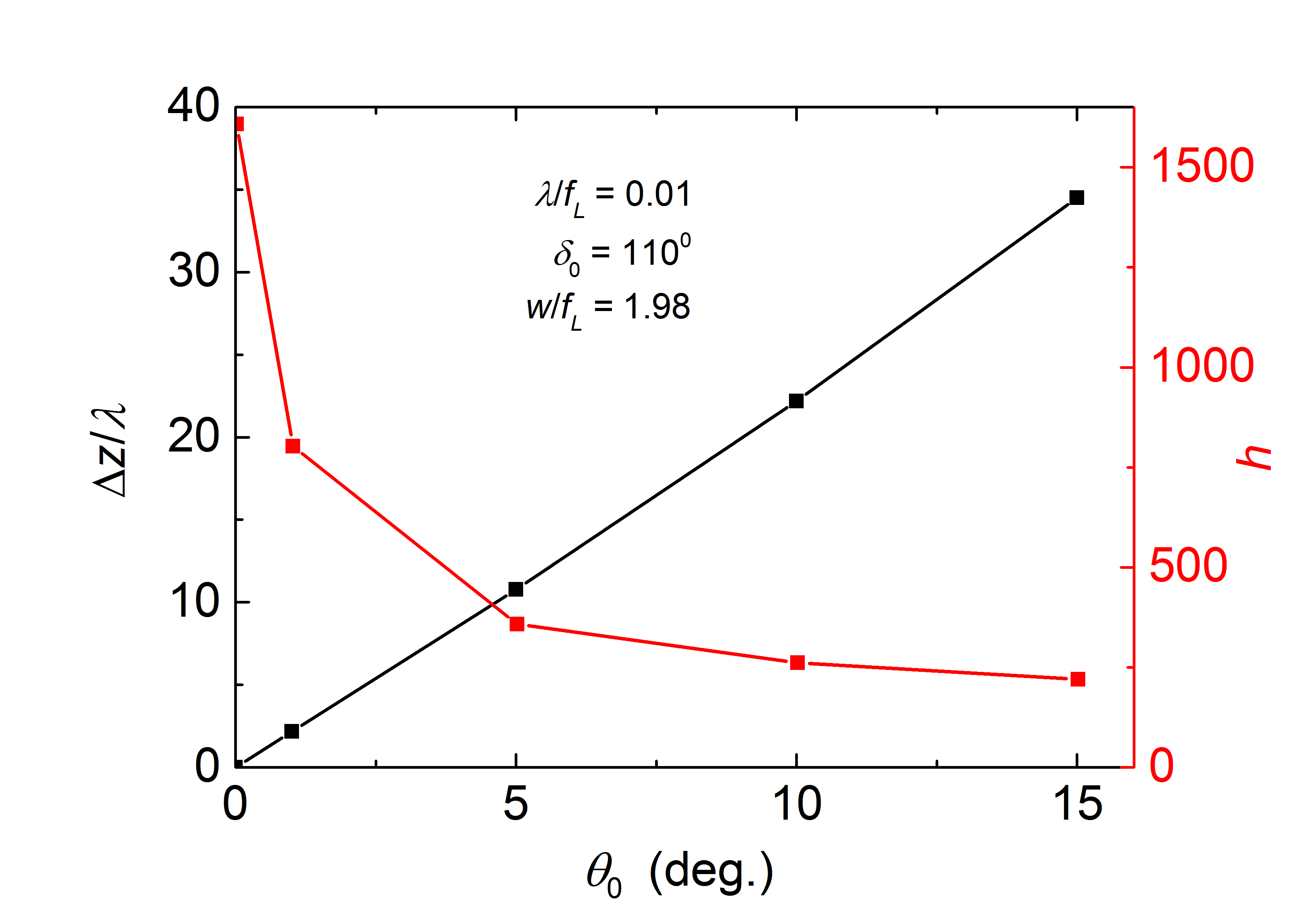}
\caption{The shift of the peak position, $\Delta z$ (normalized by the wavelength) (left scale, black line), and the field amplitude enhancement factor, $h$ (right scale, red line) versus the divergence angle $\theta_0$. \label{fig11}}
\end{figure}

It is straightforward, how to take into consideration the divergence of the incident beam during the calculations since the formulae concerning the $a$ and $b$ inputs (Eq.\ (\ref{eq20})) are well prepared for this general case, where $\theta_0$ is not negligible. Merely, in Eq.\ (\ref{eq20}) $z_w$ has to be considered according to Eq.\ (\ref{eq37}), and Eq.\ (\ref{eq33}) has to be taken into consideration when $w$ is expressed through $w_0$.

During the following investigation, the $\theta_0$-dependence was studied only for the most interesting $\vert E_z \vert$ field. The $\theta_0$ parameter was varied on the $0<\theta_0 \le 15^{\circ}$ range not exceeding the limit of validity of the theory due to the lack of terms proportional to $\mathcal{O}(\theta_0^2)$. Furthermore, occasionally the divergence of the THz beam originating from tilted-pulse-front sources falls in this range \cite{Lombosi}.

Contrary to the case of $\theta_0=0$, it is not informative enough to plot the field amplitude exactly at the focus, since the presence of the divergence results in a shift of the field maxima as it will be seen. Therefore, the $z$-dependence of $\vert E_z \vert$ was computed, and plotted in Fig.\ \ref{fig10} for $\delta_0=110^{\circ}$ ($w/f_L=1.98$) and $\lambda/f_L = 0.01$. As can be seen, if $\theta_0$ differs from zero, the symmetry of the curves breaks. The width of the curves increases rapidly with increasing $\theta_0$. Furthermore, the maxima shifts monotonously towards the positive $z$ direction as expected. This shift shows a linear dependence on $\theta_0$ with very good accuracy as can be seen in Fig.\ \ref{fig11} (black line, left scale). The widening inevitably involves a decrease in the amplitude. The field enhancement factors ($h$) are plotted versus $\theta_0$ in Fig.\ \ref{fig11} (red line, right scale). They show a significant decrease with $\theta_0$.

\begin{figure} [h]
\centering
\includegraphics[width=10.5 cm]
{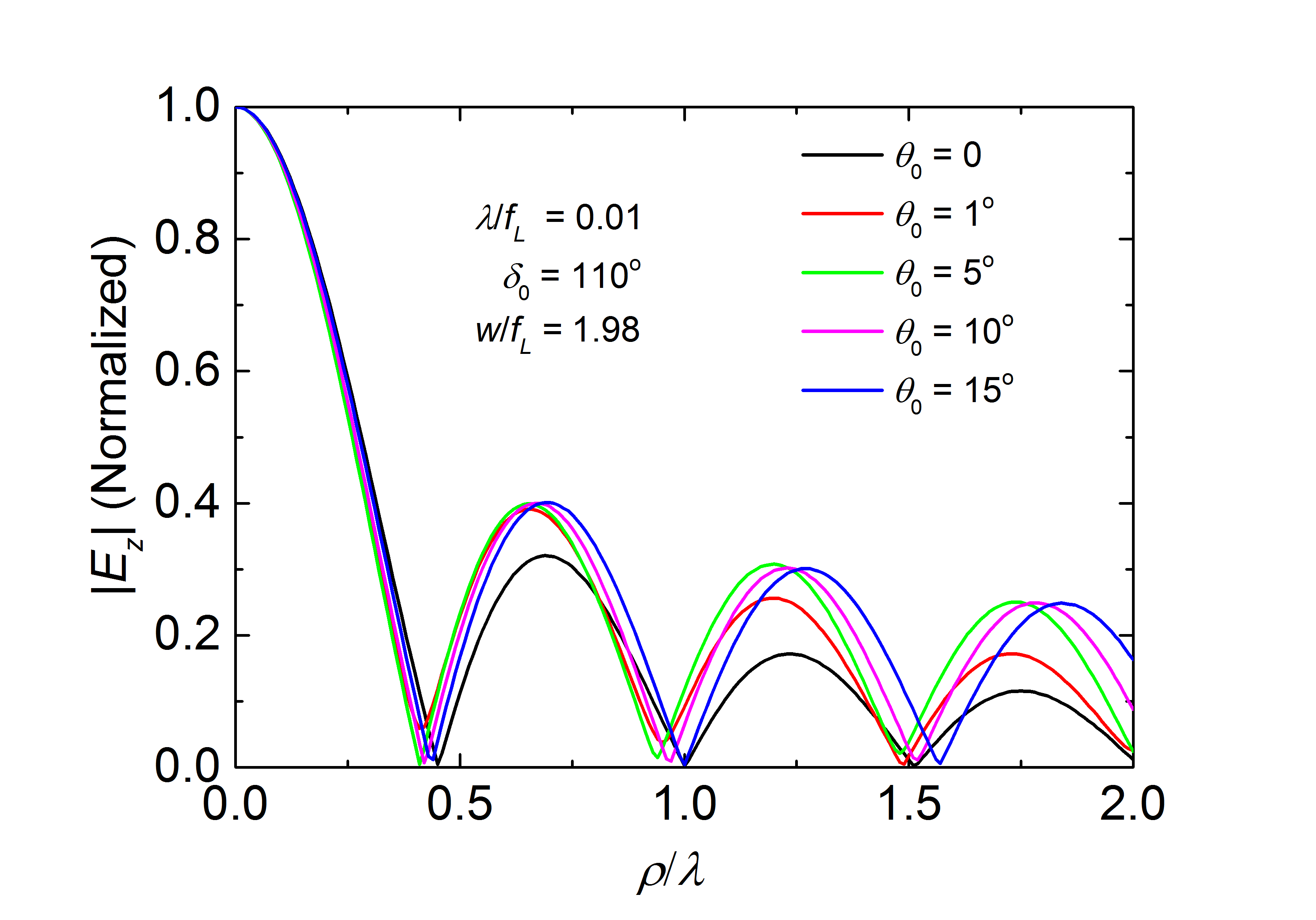}
\caption{The amplitude of the longitudinal electric field component versus the radial distance from the focus. The horizontal scale is normalized by the wavelength, and the peaks of the curves are normalized to unity. \label{fig12}}
\end{figure}

The radial distribution of $\vert E_z \vert$ was also determined at that specific $z$ position, where the amplitude of the curve (at a given $\theta_0$) had maxima. The normalized curves belonging to different $\theta_0$ do not show so significant deviation from each other along the $\rho$ radial direction (Fig.\ \ref{fig12}) as was observed for the $z$-dependence (Fig.\ \ref{fig10}). It is seen from the curves, that the amplitudes of the side maxima relative to the main maxima increase with the divergence.

\section{Conclusions}
\noindent
In this work, we presented our derived formulae concerning the electric and magnetic fields obtained upon focusing a radially polarized, monochromatic Gaussian vector beam by a parabolic mirror. The beam going to be focused can be derived rigorously from Maxwell's equations (contrary to plane waves having uniform field distribution), hence upon using the Stratton–Chu vector diffraction method an accurate, realistic picture was obtained of the field distributions. The field enhancement factor was studied in the function of the focusing angle. The results convincingly showed, that for achieving a strong longitudinal electric field a potential candidate for focusing is the ring-like paraboloid segment having a practical interest as well. In a vacuum particle accelerator, it makes possible the unobstructed entrance/exit of the particles.   Supposing the focal length/wavelength ratio to be a typical value of 100, the enhancement factor concerning the radial electric field was found to be 1610. This means, that in the terahertz frequency range longitudinal electric field component as large as $\sim$160 $\text{MV}/\text{cm}$ is available, which is ideal for particle acceleration applications. We also studied the effects of the incident beam divergence on the field distributions in the focal region. It was shown that the physical focus is shifted relative to the geometrical focus along the symmetry axis for divergent beams. This shift is linearly proportional to the beam divergence. The effect of the divergence angle on the field enhancement factor was also studied. Furthermore, it was shown that the amplitudes of the side maxima relative to the main maxima in the radial field distribution increase with the divergence.

\section*{Author contributions}
\noindent
Conceptualization, J.H., and L.P.; methodology, L.P., and Z.T.G.; software, L.P., and Z.T.G.; validation, L.P., and Z.T.G.; formal analysis, L.P., and Z.T.G.; investigation, L.P., and Z.T.G.; writing—original draft preparation, L.P., and Z.T.G.; writing—review and editing, L.P., Z.T.G. and J.H.; visualization, L.P. and Z.T.G.; supervision, L.P.; project administration, Z.T.G.; funding acquisition, J.H. and L.P.

\section*{Funding}
\noindent
The project has been supported by the Development and Innovation Fund of Hungary, financed under the TKP2021‐EGA‐17 funding scheme, by the National Research, Development and Innovation Office (2018‐1.2.1‐NKP‐2018‐00010), and by the TWAC project, which is funded by the EIC Pathfinder Open 2021 of the Horizon Europe program under grant agreement 101046504.

\section*{Data availability}
\noindent
Not applicable.

\section*{Conflicts of interest}
\noindent
The authors declare no conflict of interest.


 
\end{document}